\documentclass[10pt, conference]{IEEEtran}  
\usepackage{tikz}
\usetikzlibrary{shapes,snakes}
\usepackage{empheq}
\usepackage{graphicx}
\usepackage{epstopdf}
\usepackage{multirow}
\usepackage{tabu}
\usepackage{array}
\usepackage{url}
\usepackage{amsmath}
\usepackage{amssymb}
\usepackage{amsthm}
\usepackage{amsfonts}
\usepackage{tikz}
\usepackage{bm}
\usepackage{algorithmic}
\usepackage[ruled,vlined,linesnumbered]{algorithm2e}
\usepackage{graphicx}
\usetikzlibrary{arrows}
\usepackage{cite}
\usepackage{caption}
\usepackage{color}
\usepackage{subcaption}
\usepackage{amssymb}
\usepackage{tabulary}
\usepackage{booktabs}

\usepackage[justification=centering]{caption} 
\DeclareCaptionLabelFormat{algnonumber}{algorithm}
\tikzstyle{int}=[draw, fill=white!20, minimum size=2em]
\tikzstyle{init} = [pin edge={to-,thin,black}]

\newcommand\blfootnote[1]{%
	\begingroup
	\renewcommand\thefootnote{}\footnote{#1}%
	\addtocounter{footnote}{-1}%
	\endgroup
}

 \DeclareRobustCommand*{\IEEEauthorrefmark}[1]{%
 	\raisebox{0pt}[0pt][0pt]{\textsuperscript{\footnotesize\ensuremath{#1}}}}

\newcommand{\ve}[1]{\boldsymbol{\mathbf{#1}}}
\newcommand{\hi}{\text{High}}
\newcommand{\lo}{\text{Low}}
\newcommand{\set}[1]{\mathcal{#1}}

\IEEEoverridecommandlockouts                          

\title{
	 Dynamic Power Control for Packets with Deadlines
}

\author{
	\IEEEauthorblockN{Emmanouil Fountoulakis\IEEEauthorrefmark{\dagger\ddagger},
		Nikolaos Pappas\IEEEauthorrefmark{\dagger},
		Qi Liao\IEEEauthorrefmark{\ddagger},
		Anthony Ephremides \IEEEauthorrefmark{\dagger *},
		Vangelis Angelakis \IEEEauthorrefmark{\dagger}
	\IEEEauthorblockA{\IEEEauthorrefmark{\dagger} Department of Science and Technology, Link{\"o}ping University, Sweden}
	\IEEEauthorblockA{\IEEEauthorrefmark{\ddagger} Nokia Bell Labs, Stuttgart, Germany}
	\IEEEauthorblockA{\IEEEauthorrefmark{*} Electrical and Computer Engineering Department, University of Maryland, College Park}
	E-mails: \{emmanouil.fountoulakis, nikolaos.pappas, vangelis.angelakis\}@liu.se, etony@umd.edu\\
	\centering qi.liao@nokia-bell-labs.com}
}

\begin{document}
	
\maketitle

\thispagestyle{empty}
\pagestyle{empty}
	
\begin{abstract}

Wireless devices need to adapt their transmission power according to the fluctuating wireless channel in order to meet constraints of delay sensitive applications. In this paper, we consider delay sensitivity in the form of strict packet deadlines arriving in a transmission queue. Packets missing the deadline while in the queue are dropped from the system. We aim at minimizing the packet drop rate under average power constraints. We utilize tools from Lyapunov optimization to find an approximate solution by selecting power allocation. We evaluate the performance of the proposed algorithm and show that it achieves the same performance in terms of packet drop rate with that of the Earliest Deadline First (EDF) when the available power is sufficient. However, our algorithm outperforms EDF regarding the trade-off between packet drop rate and average power consumption.

\end{abstract}

	
\IEEEpeerreviewmaketitle 
	
\section{Introduction}
In many applications, data packets must be successfully transmitted within a particular time frame, i.e., by some deadline. If a packet is not transmitted before its deadline expiration, then, its information is considered to be useless and the packet is removed from the system \cite{hou2013packets}. This is the case for a multitude of applications, such as multimedia streaming, online gaming, and the new 5G applications such as autonomous driving that has strict round trip delay constraint. With the pervasiveness of mobile communications, such applications need to perform over wireless devices. In wireless communications, transmission errors  occur due to the fluctuating nature of the channel. Assuming perfect channel knowledge at the transmitter, the elimination of errors due to fading can be achieved by increasing the transmission power, for a given transmission rate. However, in many cases, e.g., Internet of Things (IoT), power-limited wireless devices require low average power consumption. Therefore, energy efficiency issues become very important.
\blfootnote{This work was supported in part by the European Union’s Horizon 2020 research and innovation programme under the
Marie Sk\l{}odowska-Curie grant agreements No. 643002 (ACT5G) and No. 642743 (WiVi-2020). In addition, this work was supported in part by the Center for Industrial Information Technology (CENIIT).}

Delay constrained network optimization has been extensively investigated and different optimization approaches have been applied to different scenarios, refer to \cite{SurveyDelayAwareResourceAllocation} and the references therein. For deadline-constrained scheduling, Earliest Deadline First (EDF) has been shown to be optimal in terms of number of served packets over error free (wired) channels \cite{GeorgiadisEDF}. For the case of wireless fading channels (wireless communications), the authors in \cite{OptimalTransmissionSchedulingTsitsiklis} propose an optimal scheduling scheme for single transmitter and receiver with energy constraints by using a dynamic algorithm. Similar scenarios have been studied in \cite{KumarInfocom, AnOnlineAlgorBorkar, GamalInfocom}, where dynamic programming and Markov decision theory are applied. Authors in \cite{BambosDeadlines} develop a scheduling scheme that minimizes the number of dropped packets transmitted over fading channels by using dynamic programming. In addition, they assume that the deadlines of the packets satisfy some particular requirements, i.e., the deadlines of subsequent packets depend on each other. Analytical results are provided by the authors in \cite{BambosPerfEvalu} regarding on how the power should be selected in order to approach deadlines. Authors in \cite{TTIWiOpt, PowerControlTeped, CL20182DTTI} consider deadline-constrained traffic and decide on the channel or power allocation.
In addition, authors in \cite{kam2018age} examine the impact of packet deadline on the age of information for queueing systems. In \cite{SPAWC2018}, the performance of deadline-constrained bursty traffic with retransmissions is studied.

In this paper, we develop a dynamic algorithm that finds an approximate solution to the problem of minimizing packet drop rate by optimizing power allocation under average power consumption constraints. The algorithm observes the channel conditions and the remaining deadline of the users' packets and optimizes the power allocation without knowledge of arrival packet statistics. We use Lyapunov drift and Lyapunov optimization theory to develop a dynamic algorithm. The proposed dynamic algorithm decides the power allocation at each time slot by minimizing an upper bound on the drift-plus-penalty expression. We compare the performance of our algorithm with that of EDF. EDF searches across the users the packet with the shortest expiration time and assigns to that user the appropriate power. Numerical and simulation results show that our scheduling scheme achieves the same performance in terms of packet drop rate with that of EDF when the available power is sufficient. Also, our dynamic algorithm is able to satisfy the average power constraint. On the other hand, EDF violates the average power consumption constraints when the available power is not sufficient. In addition, our dynamic algorithm offers a good trade-off between average power consumption and packet drop rate.

\section{System Model}\label{Sec:SytemModel}
\vspace{-0.2in}
We consider $N$ users transmitting packets to a single receiver over wireless fading channels. Let $\mathcal{N}\triangleq \left\{1,\dots,N\right\}$ be the set of the users in the system. Time is assumed to be slotted. Let $t \in \mathbb{Z}$ be the $t^{\text{th}}$ slot. 
\begin{table*}[t!]\caption{Notation Table.}
	\begin{center}
		\begin{tabular}{r c l l p{5cm}}
			\toprule
		     &$\mathcal{N}$    & Set of users in the system                           &  $\mathcal{P}_{i}(t)$     & Set of selectable power levels of user $i$\\
		     &$t$              & $t^{\text{th}}$ slot                                 &  $\mu_{i}(t)$             & Power allocation indicator of user $i$\\
		     &$Q_{i}(t)$       & Number of packets in queue $i$                       &  $\mathcal{P}(t)$         & Set of power constraints for $\mathbf{p}(t)$\\
		     &$\pi_{i}$          & Packet arrival probability of user $i$                      &  $D_{i}(t)$               & Packet drop indicator of user $i$\\
		     &$\alpha_{i}(t)$  & Packet arrival indicator of user $i$                 &  $\overline{D}_{i}$       & Packet drop rate of user $i$\\
 		     &$m_{i}$          & Deadline of packet of user $i$                       &  $\overline{p}_{i}$       & Average power consumption of user $i$\\
		     &$d_{i}(t)$       & Number of slots left before the deadline of user $i$ & $X_{i}(t)$                & Length of virtual queue of user $i$\\
		     &$\mathbf{S}(t)$  & Channel states                                       & $L(\cdot)$                & Quadratic Lyapunov function \\
		     & $\mathbf{p}(t)$ & Power allocation vector                              & $\Delta(L(\cdot))$        & Lyapunov drift\\
 		     & $\gamma_{i}$    & Allowed average power consumption for user $i$       & $\bm{\alpha}(t)$             & Packet arrival indicator vector\\
			\bottomrule
		\end{tabular}
	\end{center}
	\label{tab:TableOfNotationForMyResearch}
\end{table*}We consider the users to be synchronized and at most one user can transmit at each time slot.
Each user $i$, where $i \in \mathcal{N}$, is associated with a queue where the packets are held or dropped. Let $Q_{i}(t)$ be the number of packets in queue $i$ in the $t^{\text{th}}$ slot. Each user $i$ generates a packet with a probability $\pi_{i}$ at each time slot $t$. Let $\bm{\alpha}(t)\triangleq \left\{\alpha_{i}(t)\right\}_{i\in \mathcal{N}}$, where $\alpha_{i}(t) \in \left\{0,1\right\}$, represent the packet arrival process for each user $i$ in the $t^{\text{th}}$ slot. The random variables of packet arrival process are independent and identically  distributed (i.i.d.). Furthermore, we assume that at most one packet can be transmitted at each time slot and no collisions are allowed.

Each packet that arrives in a queue has a deadline by which it must be transmitted. Otherwise, it is dropped and removed from the system. For simplicity, the  deadlines of the packets in the same queue are assumed to be the same. However, deadlines of different queues may vary. We denote the packet deadline of the $i^{\text{th}}$ queue with $m_{i} \in \mathbb{Z}_{+}\text{, }\forall i \in \mathcal{N}$. We assume that in each queue, packets are served in the order that they arrive following the First In First Out (FIFO) discipline. Let $d_{i}(t)$ be the number of slots left in the $t^{\text{th}}$ slot before the packet that is at the head of queue $i$ expires.

We assume that the channel state at the beginning of each time slot is known. The channel state remains constant within one slot but it changes from slot to slot. Let $\mathbf{S}(t)\triangleq \left\{S_{i}(t)\right\}_{i\in \mathcal{N}}$ represent the channel state for each user $i$ during slot $t$. We assume that the channel can be either in ``Bad" state (deep fading) or in ``Good" state (mild fading). The possible channel states of each user $i$ are described by the set $\mathcal{S}\triangleq\left\{\text{B},\text{G}\right\}$, and $S_{i}(t)\in \mathcal{S}$, $\forall i \in \mathcal{N}$. For simplicity, we assume that the random variables of the channel process $\mathbf{S}(t)$ are i.i.d. from one slot to the next. 

Let $\ve{p}(t)\triangleq \left[p_{1}(t),\ldots,p_{N}(t)\right]$ denote the power allocation vector in the $t^{\text{th}}$ slot. We consider a set of discrete power levels $\left\{0, P^{(\lo)}, P^{(\hi)}\right\}$. We assume that $P^{(\hi)}$ is needed for a packet to be successfully transmitted under ``Bad" channel condition, and $P^{(\lo)}$  under ``Good" channel condition. At each time slot, the set of selectable power levels $\set{P}_{i}(t)$ for each user is conditioned on the channel state $S_i(t)$. For example, if the current channel state is ``Bad", then $P^{\text{(Low)}}$ cannot be selected. Thus, we have 
\begin{align}\label{IndividualPowSet}
     p_{i}(t) \in \begin{cases}
    \left\{0, P^{\text{(High)}}\right\}\text{, if } S_i(t) = \text{B}\\
     \left\{0, P^{\text{(Low)}}\right\}\text{, if } S_i(t) = \text{G}
    \end{cases}\text{, }\forall i\in\set{N}\text{.}
\end{align}
Let $\mu_{i}(t)$ be the power allocation, or packet serving, indicator for the user $i$ in the $t^{\text{th}}$ slot, we have
\begin{align}\label{AllocIndic}
   \mu_{i}(t) \triangleq \begin{cases}
    1\text{, if } p_i(t)>0\\
    0\text{, otherwise}
    \end{cases}\text{, }\forall i\in\set{N}\text{.}
\end{align}
%
At most one packet can be transmitted in a timeslot $t$, i.e., the vector $\mathbf{p}(t)$ has at most one non-zero element. The set of power constraints for $\ve{p}(t)$ is then defined by 
\begin{equation}\label{eqn:GeneralPowSet}
\set{P}(t)\triangleq \left\{\ve{p}(t): \sum_{i=1}^N \mathbf{1}_{\{\mu_i(t)=1\}}\leq 1\right\}\text{,}
\end{equation}
where $\mathbf{1}_{\{\cdot\}}$ denotes the indicator function. 

In our system, a packet is dropped if its deadline has expired. Since the queue follows FIFO discipline, a packet is dropped under the following conditions: 1) it is at the head of the queue; 2) the remaining number of the slots to serve the packet is $1$; and 3) power is not assigned to $i$ at the current slot. Let $D_{i}(t)$ be the indicator of the packet drop for user $i$ at time $t$.
The queue evolution is described as
\begin{align}\label{QueueEvolution}
	Q_{i}(t+1) \triangleq \max \left[Q_{i}(t)- \mu_{i}(t),0 \right] 
	+ \alpha_{i}(t)-D_{i}(t)\text{, } \forall i \in \mathcal{N}\text{.}
\end{align}

Furthermore, we assume that $Q_{i}(0)=0$, and $D_{i}(0)=0$, $\forall i \in \mathcal{N}$.
Let
\begin{align}\label{DropRate}
    \overline{D}_{i} & \triangleq \lim\limits_{t\rightarrow\infty}  \overline{D}_{i}(t)\text{, } \forall i \in \mathcal{N}\text{,}\\
    \overline{p}_{i} & \triangleq \lim\limits_{t\rightarrow \infty} \overline{p}_{i}(t)\text{, }
    \forall i \in \mathcal{N}\text{,}
\end{align}
respectively denote the packet drop rate and the average power consumption, where $\overline{D}_{i}(t)=\frac{1}{t}\sum\limits_{\tau=0}^{t-1} D_{i}(\tau)$ and $\overline{p}_{i}(t)=\frac{1}{t}\sum\limits_{\tau=0}^{t-1} p_{i}(\tau)$. The packet drop rate represents the average number of dropped packets per time slot. The average power consumption represents the average of transmit power over all time slots. These metrics are connected and we will show in the following sections how the average power consumption affects the packet drop rate.

\section{Problem Formulation}\label{sec:formulation}
We desire a scheduling scheme that offers fairness among users when minimizing their packet drop rate under average power constraints. Furthermore, we are interested in the trade-off between packet drop rate and time average power consumption.
%
%
To this end, we present the following problem
\begin{subequations}
	\begin{align}\label{Opt.Probl}
	\min\limits_{\bm{p}(t)} \quad & \sum\limits_{i=1}^{N} \overline{D}_{i}\\\label{PowerConstr}
	\text{s.~t.} \quad & \overline{p}_{i}\leq \gamma_{i}\text{, } \forall i \in  \mathcal{N}\text{,}\\\label{PowerAlloc}
	\quad & \bm{p}(t) \in \mathcal{P}(t)\text{,}
	\end{align}
\end{subequations}
where $\gamma_{i} \in \left[0,P^{\text{(High)}}\right]$ indicates the allowed average power consumption. The constraint in (\ref{PowerConstr}) ensures that average power consumption of each user $i$ remains below $\gamma_{i}$ power units.

The formulation above represents our intended goal which is the minimization of the packet drop rate. However, the objective function in (\ref{Opt.Probl}) has a basic disadvantage that makes the solution approach non-trivial. The decision variable, $\mathbf{p}(t)$ (power allocation), is optimized slot-by-slot for minimization of the objective function that is defined over infinite horizon. We have to cope with one critical point: We do not have any knowledge about the future states of the channel and packet arrival in the system. Therefore, we are not able to predict the values of the objective function in the future slots in order to decide on the power allocation that minimizes the cost. We aim to design a function whose future values are affected by the current decision and the remaining expiration time of the packets. 
To this end, we introduce a function incorporating the relative difference between the packet deadline $m_{i}$ and the number of remaining future slots $(d_{i}(t)-1)$ before its expiration as described below
\begin{equation}\label{NewFunction}
f_{i}(t) \triangleq \frac{m_{i}-(d_{i}(t)-1)}{m_{i}} \mathbf{1}_{\left\{\mu_{i}(t)=0\right\}}.
\end{equation}
The function in (\ref{NewFunction}) takes its extreme value $f_{i}(t)=0$ when a packet of user $i$ is served, or $f_{i}(t)=1$ when a packet of user $i$ is dropped. Therefore, that function takes the same values with those of (\ref{DropRate}) in the extreme cases. In addition, the function in (\ref{NewFunction}) assigns the cost according to the remaining time of a packet to expire  in the intermediate states, i.e., when a packet is waiting in the queue. The cost increases when there is less time left for serving the packet with respect to the defined deadline. The time average of $f_{i}(t)$ is 
\begin{align}\label{eqn:NewObj}
\overline{f_{i}} \triangleq \lim\limits_{t\rightarrow \infty} \overline{f}_{i}(t)\text{,}
\end{align}
where $\overline{f}_{i}(t)\triangleq \frac{1}{t} \sum\limits_{\tau=0}^{t-1} f_i(\tau)$.
Finally, we formulate a minimization problem by using \eqref{eqn:NewObj} as shown below
\begin{subequations}\label{Eq.Probl}
	\begin{align}  
	\min\limits_{\bm{p}(t)} \quad & \sum\limits_{i=1}^{N} \overline{f}_{i}\\\label{PowerConstr2}
	\text{s.~t.} \quad & \overline{p}_{i}\leq \gamma_{i} \text{, } \forall i \in \mathcal{N}\text{,}\\\label{PowerAlloc2}
	\quad & \bm{p}(t) \in \mathcal{P}(t)\text{.}
	\end{align}
\end{subequations}

\section{Proposed Approximate Solution} 
The problem in (\ref{Eq.Probl}) includes time average constraints. In order to satisfy these constraints, we aim to develop a policy that uses techniques different from classic optimization methods based on static and deterministic models. For example, policies that select power less than $\gamma_{i}$ at every time slot ensures that constraint (\ref{PowerConstr2}) is satisfied. However, this kind of policies decrease the degrees of freedom of power selection. In Table \ref{Table:Example}, we provide an illustrative example with one user. We consider that $P^{\text{(Low)}}=1$ power units, and $P^{\text{(High)}}=2$ power units. In this example, the average power   consumption must be less than or equal to $1.5$ power units, i.e., $\gamma=1.5$ power units (subscripts are omitted for simplicity). We compare the performance of two policies $\omega_{1}$ and $\omega_{2}$. Policy $\omega_{1}$ selects power less than $\gamma$ power units at every time slot in order to restrict the average power consumption below $1.5$ power units. On the other hand, $\omega_{2}$ allows power selection greater than $1.5$ power units for each time slot. We observe that policy $\omega_{2}$ achieves better performance than $\omega_{1}$ by satisfying the power constraint. This motivates us to look for a more efficient way to satisfy the average power consumption constraint.
\begin{table}[t!]
	\centering
	\begin{tabular}{|c|c|c|c|c|}
		\cline{3-5}
		\multicolumn{2}{c|}{\multirow{2}{*}{}} & $t=1$  & $t=2$ & $t=3$ \\
		\multicolumn{2}{c|}{\multirow{2}{*}{}} & $S(t) = \text{B}$ & $S(t) = \text{G}$ & $S(t) = \text{G}$    \\
		\hline
		\multirow{2}{*}{$d(t)$} & $\omega_{1}$ &   $1$ & $2$ & empty queue   \\
		&$\omega_{2}$ &   $1$ & $2$ & $1$   \\
		\hline
		\multirow{2}{*}{$p(t)$} & $\omega_{1}$ & 0 & 1 & 0  \\
		& $\omega_{2}$ & 2 & 0 & 1  \\
		\hline
	\end{tabular}	
	
	\vspace{2mm}
	
	\centering
	\begin{tabular}{|c|c|c|c|}
		\cline{3-4}
		\multicolumn{2}{c|}{\multirow{2}{*}{}} & $\overline{p}(t)$  & drop packets  \\
		\hline
		\multirow{2}{*}{$p(t)$} & $\omega_{1}$ &       1         &      1         \\
		& $\omega_{2}$  &       1.5       &      0     \\
		\hline
	\end{tabular}
	\caption{Example showing the gain achieved by deciding different power allocations.}
	\label{Table:Example}
\end{table}

We apply the technique developed in \cite{NeelyEnergyOptimalControl} and further discussed in \cite{NeelyBook} and \cite{TassiulasNeelyNow} in order to develop a policy that ensures that the constraint in (\ref{PowerConstr2}) is satisfied. Each inequality constraint in (\ref{PowerConstr2}) is mapped to a virtual queue. We show below that the power constraint problem is transformed into a queue stability problem. 

Let $\left\{X_{i}(t)\right\}_{i\in \mathcal{N}}$ be the virtual queues associated with constraint (\ref{PowerConstr2}). We update each virtual queue $i$ at each time slot $t$ as
\begin{align}
X_{i}(t+1) \triangleq \max\left[X_{i}(t)-\gamma_{i}, 0\right] + p_{i}(t)\text{.}
\end{align}
Process $X_{i}(t)$ can be viewed as a queue  with ``arrivals" $p_{i}(t)$ and ``service rate" $\gamma_{i}$.

Before describing the motivation behind the mapping of power constraints to virtual queues, let us recall one basic theorem that comes from the general theory of stability of stochastic processes \cite{MarkovChains}. Consider a system with $K$ queues. The number of unfinished jobs of queue $i$ are denoted by $q_{i}(t)$ and
$\mathbf{q}(t) = \left\{q_{i}(t)\right\}_{k=1}^{K}$. The Lyapunov function and the Lyapunov drift are denoted by $L(\mathbf{q}(t))$ and
$\Delta(L(\mathbf{q}(t))) \triangleq E\left\{L(\mathbf{q}(t+1))-L(\mathbf{q}(t)) | \mathbf{q}(t)\right\}$ respectively \cite{MarkovChains}.
Before describing the Lyapunov Drift theorem, let us recall the definition of the Lyapunov function \cite{MarkovChains}.

\textit{Definition 1 (Lyapunov function):} A function $L: \mathbb{R}^{K}\rightarrow \mathbb{R}$ is said to be a Lyapunov function if it has the following properties
\begin{itemize}
	\item $L(\mathbf{x})\geq 0\text{, } \forall \mathbf{x} \in \mathbb{R}^{K}$,
	\item It is non-decreasing in any of its arguments,
	\item $L(\mathbf{x}) \rightarrow + \infty$, as $||\mathbf{x}||\rightarrow + \infty$.
\end{itemize}

\textit{Theorem 1} \textit{(Lyapunov Drift):} If there are positive values $B$, $\epsilon$ such that for all time slots $t$ we have
$\Delta (L(\mathbf{q}(t)) \leq B - \epsilon \sum\limits_{k=1}^{K} q_{n}(t)\text{,}$
then the system $\mathbf{q}(t)$ is \textit{strongly} stable. 

In below, we show that the power constraint problem is transformed into a queue stability problem. Then, we develop a dynamic algorithm that satisfies \text{Theorem 1} in order to achieve stability.

\textit{Theorem 2}: If $X_{i}(t)$ is \textit{rate stable}\footnote{A discrete time process $Q(t)$ is \textit{rate stable} if $\lim\limits_{t\rightarrow \infty}\frac{Q(t)}{t}=0$ with probability $1$ \cite{NeelyBook}.}, then the constraint in (\ref{PowerConstr2}) is satisfied.
\begin{proof}
	Using the basic sample property \cite[Lemma 2.1, Chapter 2]{NeelyBook}, we have
	\begin{align}
	\frac{X_{i}(t)}{t}-\frac{X_{i}(0)}{t} \geq  \frac{1}{t} \sum\limits_{\tau = 0} ^{t-1} p_{i}(\tau) - \frac{1}{t}\sum\limits_{\tau = 0} ^{t-1} \gamma_{i}\text{.}
	\end{align}
	Therefore, if $X_{i}(t)$ is rate stable, so that $\frac{X_{i}(t)}{t}\rightarrow 0\text{, } \forall i$, with probability 1, then constraint (\ref{PowerConstr2}) is satisfied with probability $1$ \cite{QueueStabilityNeely}.
\end{proof}
Note that strong stability implies all of the other forms of stability \cite[Chapter 2]{NeelyBook} including the rate stability. Therefore, the problem is transformed into a queue stability problem.
In order to stabilize the virtual queues $X_{i}(t)\text{, } \forall i \in \mathcal{N}$, we first define our Lyapunov function as
\begin{align}
    L(\mathbf{X}(t)) \triangleq \frac{1}{2} \sum\limits_{i=1}^{N} X_{i}(t)^2\text{,}
\end{align}
where $\mathbf{X}(t)=\left\{X_{i}(t) \right\}_{i\in \mathcal{N}}$ and the Lyapunov drift as
\begin{align}
    \Delta(\mathbf{X}(t)) \triangleq \mathbb{E}\left\{ L(\mathbf{X}(t+1)) - L(\mathbf{X}(t)) | \mathbf{X}(t)  \right\}.
\end{align}
The above conditional expectation is with respect to the random channel states and the arrivals. 

To minimize the time average of the desired cost $f_{i}(t)$ while stabilizing the virtual queues $X_{i}(t)$, $\forall i \in \mathcal{N}$, we use the \textit{drift-plus-penalty} minimization approach introduced in \cite{TassiulasNeelyNow}. The approach seeks to minimize an upper bound on the following drift-plus-penalty expression at every slot $t$:
\begin{align}
   \Delta(\mathbf{X}(t)) + V\sum\limits_{i \in \mathcal{N}} \mathbb{E}\left\{f_{i}(t)|\mathbf{X}(t)\right\}\text{,}
\end{align}
where $V>0$ is an ``importance" weight to scale the penalty.

We derive an upper bound for the drift by using the fact $(\max\left[Q-b,0\right] + A)^2 \leq Q^2 + A^2 + b^2 +2Q(A-b)$ as shown below
\begin{align}\label{ineq:powerqueue} 
   X_{i}(t+1)^2  \leq X_{i}(t)^2 + p^2_{i}(t)
   + 2X_{i}(t)(p_{i}(t)-\gamma_{i}) + \gamma_{i}^2\text{.}
\end{align}
Taking the sum over all the queues in (\ref{ineq:powerqueue}) we have
\begin{small}
	\begin{align}\label{Ineq:BDrif}\nonumber
	\sum\limits_{i=1}^{N}\frac{X_{i}(t+1)^2}{2} - \sum\limits_{i=1}^{N}\frac{X_{i}(t)^2}{2} & \leq \sum\limits_{i=1}^{N} \frac{X_{i}(t)^2+p_{i}(t)^2+\gamma_{i}^2}{2}\\ & + \sum_{i=1}^{N} X_{i}(t) (p_{i}(t)-\gamma_{i})\text{.}
	\end{align}
\end{small}

Taking the expectations in (\ref{Ineq:BDrif}), we have
\begin{align}\label{ineq:driftbound}
    \Delta(\mathbf{X}(t)) \leq B + \sum\limits_{i=1}^{N}\ X_{i}(t) \mathbb{E}\left\{y_{i}(t)| \mathbf{X}(t)\right\}\text{,}
\end{align}
where $y_{i}(t) = p_{i}(t) -\gamma_{i}$, and $B$ is constant,
\begin{align}
    B\geq \frac{1}{2} \sum\limits_{i=1}^{N} \mathbb{E}\left\{X_{i}(t)^2 + p_{i}(t)^2 + \gamma_{i}^2 | \mathbf{X}(t)\right\}\text{.}
\end{align}
Therefore, an upper bound for the drift plus penalty expression is
\begin{small} 
	\begin{align}\label{Ineq:UpperBound}\nonumber
	&\Delta(\mathbf{X}(t)) + V\sum_{i=1}^{N} \mathbb{E}\left\{f_{i}(t) | \mathbf{X}(t)\right\}\\ &\leq B
	+ \sum\limits_{i=1}^{N} X_{i}(t) \mathbb{E}\left\{y_{i}(t)| \mathbf{X}(t)\right\} + V\sum\limits_{i=1}^{N} \mathbb{E}\left\{f_{i}(t)|\mathbf{X}(t)\right\}\text{.}
	\end{align}
\end{small}
\subsection{Min-Drift-Plus-Penalty Algorithm}
Note that the power allocation decision on slot $t$ affects only the last two terms in (\ref{Ineq:UpperBound}). The proposed algorithm observes the virtual queue backlogs $\mathbf{X}(t)$ and the channel states and makes a control action to minimize the following expression
\begin{align}
\sum\limits_{i=1}^{N}X_{i}(t) \mathbb{E}\left\{y_{i}(t)| \mathbf{X}(t)\right\}+ V\sum\limits_{i=1}^{N} \mathbb{E}\left\{f_{i}(t)|\mathbf{X}(t)\right\}\text{.}
\end{align}
 The algorithm decides the power allocation by solving the following optimization problem at each time slot
\begin{subequations}\label{Opt:MinDrift}
	\begin{align}
	\min\limits_{\bm{p}(t)} \quad & V\sum\limits_{i=1}^{N} f_{i}(t)  +  \sum\limits_{i=1}^{N}X_{i}(t)y_{i}(t)\\
	\quad & \bm{p}(t) \in \mathcal{P}(t)\text{.}
	\end{align}
\end{subequations}

In the following we show that the optimal solution to problem (\ref{Opt:MinDrift}) minimizes the upper bound of the drift-plus-penalty
expression given in the right-hand-side of (\ref{Ineq:UpperBound}). 
Let $\mathbf{p}(t)$ represent any, possibly randomized, power allocation decision made at slot $t$. Suppose  that $\mathbf{p}^{*}(t)$ is the optimal solution to problem  (\ref{Opt:MinDrift}), and under action $\mathbf{p}^{*}(t)$ the value of $f_{i}(t)$ yields $f_{i}^{*}(t)$, and that of $y_{i}(t)$, $y^{*}(t)$, we have 
\begin{align}\label{Ineq:OptBound}
    V\sum\limits_{i=1}^{N}  f^{*}_{i}(t)  +  \sum\limits_{i=1}^{N}X_{i}(t)y^{*}_{i}(t) \leq V\sum\limits_{i=1}^{N} f_{i}(t)  +  \sum\limits_{i=1}^{N} X_{i}(t)y_{i}(t)\text{.}
\end{align}
Taking the conditional expectations of (\ref{Ineq:OptBound}), we have
\begin{align}\nonumber
 V\sum\limits_{i=1}^{N} \mathbb{E}\left\{f^{*}_{i}(t)|\mathbf{X}(t)\right\}+
 \sum\limits_{i=1}^{N} X_{i}(t) \mathbb{E}\left\{y^{*}_{i}(t)| \mathbf{X}(t)\right\}\leq \\
  V\sum\limits_{i=1}^{N} \mathbb{E}\left\{f_{i}(t)|\mathbf{X}(t)\right\}+
 \sum\limits_{i=1}^{N} X_{i}(t) \mathbb{E}\left\{y_{i}(t)| \mathbf{X}(t)\right\}\text{.}
\end{align}
In view of the above, it is concluded that the optimal solution to problem  (\ref{Opt:MinDrift}) minimizes the upper bound given in the right-hand-side of  (\ref{Ineq:UpperBound}). Note that the solution we provide is an approximate solution because we minimize an upper bound of the drift defined in (\ref{Ineq:UpperBound}). Furthermore, we find an approximate solution of the problem in (\ref{Eq.Probl}) by solving a snapshot problem (\ref{Opt:MinDrift}) for a particular time slot $t$. 

We summarize the steps of the proposed dynamic control algorithm to solve problem (\ref{Eq.Probl}) in Algorithm \ref{alg:MyAlgorithm}, named dynamic power allocation (DPA) algorithm. DPA uses exhaustive search that solves the problem in (\ref{Opt:MinDrift}).

\setlength{\textfloatsep}{0pt}
\begin{algorithm}[!h]
	\scriptsize\nonumber
	\caption{DPA}\label{alg:MyAlgorithm}
	Input constant $V$,
	Initialization $X_{i}(0)=0\text{, } \gamma_{i}\text{, }\forall i \in \mathcal{N}$\\
	\For{$t=1:\ldots$}{
		$MinObj\leftarrow \infty$\\
		\For{$i\in \mathcal{N}$}{
			$p_{i}(t) \in \mathcal{P}(t)$, 
			Calculate $f_{j}(t)\text{, } \forall j \in \mathcal{N}$\\
			$Obj\leftarrow V\sum\limits_{j=1}^{N}f_{j}(t) + \sum\limits_{j=1}^{N}X_{j}(t)y_{j}(t)$\\
		    \If{MinObj$>$Obj}
		    {$\mathbf{p}'(t)\leftarrow \mathbf{p}(t)$\\
		     $MinObj\leftarrow Obj$}
	    }
       $\mathbf{p}(t)\leftarrow \mathbf{p}'(t)$\\
        $X_{j}(t+1)\leftarrow \max\left[X_{j}(t)-\gamma_{j}, 0\right] + p_{j}(t)\text{, } \forall j \in \mathcal{N}$
	}
\end{algorithm}

 In step $1$, we initialize $V$ and the length of virtual queues. We calculate the value of the objective function for each possible value of vector $\mathbf{p}(t)$ as shown in steps $5$--$6$. In step $7$, we compare each possible value of the objective function (for different power allocations) and keep the corresponding power allocation in vector $\mathbf{p}'(t)$ as shown in step $8$. 
We decide power allocation as shown in step $10$. 
The complexity of DPA is $\mathcal{O}(N^{2})$. 	

\section{Numerical and Simulation Results}
In this section, we compare the performance of DPA with that of earlier deadline first (EDF) algorithm. Recall that EDF finds across the users the packet with the shortest remaining expiration time and it assigns to its user the appropriate power according to the channel conditions.
We compare the performance of the algorithms in terms of packet drop rate and average power consumption and we show the trade-off between them. Additionally, we provide results showing the performance of our algorithm for different values of $V$ and how they affect the average power consumption.

In the simulation setup, the probability a channel to be in ``Bad" and ``Good" state is $0.6$ and $0.4$, respectively.
Also, we consider that the arrival process for each user $i$ is an i.i.d. Bernoulli process with probability $\overline{\lambda}_{i}$. In addition, we consider that $P^{\text{(Low)}}=1$, and $P^{\text{(High)}}=2$. The deadlines are $m_{1}=m_{2}=5$ time slots.
\begin{figure}[t!]
	\begin{subfigure}{.5\textwidth}
		\centering
		\includegraphics[scale=0.45]{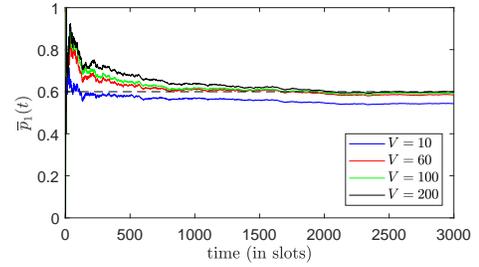}
		\caption{Average power consumption.}
		\label{Fig:convergence}
	\end{subfigure}%
	
	\centering
	\begin{subfigure}{.5\textwidth}
		\centering
		\includegraphics[scale=0.45]{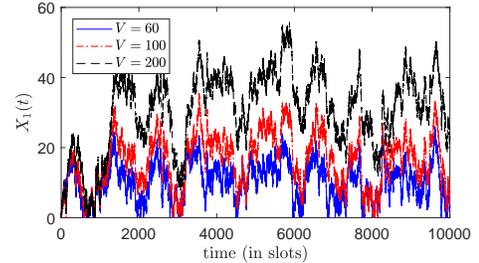}
		\caption{Virtual queue evolution.}
		\label{Fig:Xevolution}
	\end{subfigure}
	
	\centering
	\begin{subfigure}{.5\textwidth}
		\centering
		\includegraphics[scale=0.45]{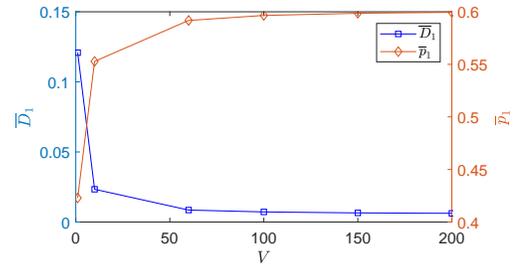}
		\caption{Tradeoff between packet drop rate and average power consumption.}
		\label{Fig:DropVvalues}
	\end{subfigure}
	\vspace{-0.05in}
	\caption{DPA performance depending on $V$. $\gamma_{1}=\gamma_{2}=0.6$, $\overline{\lambda}_{1}=\overline{\lambda}_{2}=0.4$.}
	\label{Fig:TradeOffDropsPower}
\end{figure}

Fig. \ref{Fig:TradeOffDropsPower} depicts how different values of $V$ affect the packet drop rate and the average power consumption of user $1$. We observe that the larger the value of $V$ the slower the convergence of the algorithm in terms of power rate consumption constraint. However, it is shown in Fig. \ref{Fig:convergence} that even for large values of $V$, DPA is able to keep the power rate consumption below $\gamma_{i}$ and, therefore to satisfy the power consumption constraint. For large values of $V$, DPA allows virtual queue backlogs to take large values as shown in Fig. \ref{Fig:Xevolution}. The reason why the backlogs of the virtual queues increase is because the dominant term of the objective function is the one that includes $V$. However, as the time passes by the virtual queue backlog increases and dominates the penalty term that includes $V$. Thus, DPA allocates lower power in order to decrease the virtual queue backlog and stabilizes it as shown in Fig. \ref{Fig:Xevolution}. In Fig. \ref{Fig:DropVvalues}, we provide results for different values of $V$. We show the trade-off between the average power consumption and the packet drop rate. As expected, the average power consumption increases with increasing value  of $V$. However, the average power consumption is always below $0.6$.

\begin{figure}[t!]
	\centering
	\includegraphics[scale=0.45]{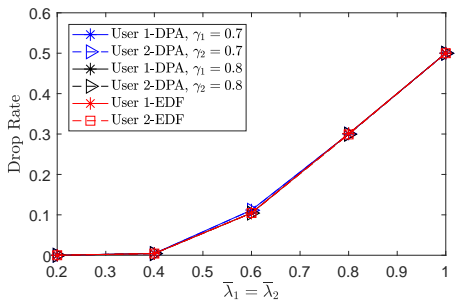}
	\vspace{-0.05in}
	\caption{Packet drop rate for EDF and DPA for different values of $\overline{\lambda}_{i}$ and $\gamma_{i}$.}
	\label{Fig:EDFvsDPADropRate}
\end{figure}

\begin{figure}[t!]
	\centering
	\includegraphics[scale=0.45]{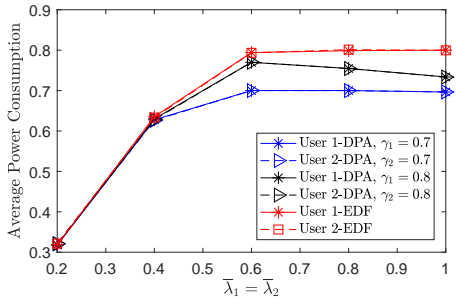}
	\vspace{-0.05in}
	\caption{Average power consumption for EDF and DPA for different values of $\overline{\lambda}_{i}$ and $\gamma_{i}$.}
	\label{Fig:EDFvsDPAPowerRate}
\end{figure}

Values of $V$ that are larger than $60$ do not affect significantly the packet drop rate. Thus we present the rest of the simulation results for $V=60$. In Fig. \ref{Fig:EDFvsDPADropRate} and Fig. \ref{Fig:EDFvsDPAPowerRate}, we compare the performance of the two algorithms in terms of packet drop rate and average power consumption. Note that EDF does not take into account the average power consumption of each user. Therefore, for some values of $\gamma_{i}$, EDF algorithm violates the average power constraints. For example, we see in Fig. \ref{Fig:EDFvsDPAPowerRate} that EDF algorithm violates the average power constraints for $\gamma_{1}=\gamma_{2}=0.7$. The performance of DPA in terms of packet drop rate is very close to that of EDF. However, we observe in Fig. \ref{Fig:EDFvsDPAPowerRate}, the average power consumption of DPA is lower than that of EDF by $0.1$ power units. For $\gamma_{1}=\gamma_{2}=0.8$, we observe that our algorithm has the same performance in terms of packet drop rate with that of EDF. However, in Fig. \ref{Fig:EDFvsDPAPowerRate}, we see that the average power consumption of DPA decreases when the traffic arrival exceeds a sufficiently large value, i.e., for $\overline{\lambda}_{1}=\overline{\lambda}_{2}>0.6$. The reason why the average power consumption decreases is because that for large values of $\lambda_{i}$, the scheduler has often to cope with users having packets with one time slot left before their expiration. Thus, it selects to assign power to the user who has the best channel condition and drops the packet of the user with the worst channel condition.

Overall, we observe that our algorithm performs as the EDF algorithm when the power limit  is sufficiently high. Furthermore, the proposed algorithm is able to satisfy the average power constraints of the users and offer a good trade-off between packet drop rate and average power consumption as shown in Fig. \ref{Fig:EDFvsDPADropRate} and Fig. \ref{Fig:EDFvsDPAPowerRate}.

\section{Conclusions}
\vspace{-0.1in}
In this paper, we propose a dynamic algorithm that decides power allocation at each time slot by minimizing an objective function. The proposed algorithm is based on Lyapunov optimization theory. We evaluate the performance of the proposed algorithm through simulations and compare it with EDF. We observe that our proposed algorithm has the same performance with EDF in terms of packet drop rate when the available power is sufficient. Furthermore, the proposed scheduling scheme can handle packets with deadlines and control the transmission power of the devices. Since we have systems with mobile devices and therefore, limited available power, it is important to develop a dynamic algorithm that satisfies the average power constraints of each user.

\vspace{-0.1in}
\bibliographystyle{IEEEtran}
\bibliography{MyBib1}

\end{document}